\documentclass[a4paper,UKenglish,cleveref, autoref, thm-restate,authorcolumns]{lipics-v2021}

\title{From Verification to Causality-based Explications} 

\author{Christel Baier}{Technische Universität Dresden, Germany }{christel.baier@tu-dresden.de}{https://orcid.org/0000-0002-5321-9343}{}

\author{Clemens Dubslaff}{Technische Universität Dresden, Germany }{clemens.dubslaff@tu-dresden.de}{https://orcid.org/0000-0001-5718-8276}{}

\author{Florian Funke}{Technische Universität Dresden, Germany }{florian.funke@tu-dresden.de}{https://orcid.org/0000-0001-7301-1550}{}

\author{Simon Jantsch}{Technische Universität Dresden, Germany }{simon.jantsch@tu-dresden.de}{https://orcid.org/0000-0003-1692-2408}{}

\author{Rupak Majumdar}{MPI-SWS, Kaiserslautern, Germany}{rupak@mpi-sws.org}{https://orcid.org/0000-0003-2136-0542}{}

\author{Jakob Piribauer}{Technische Universität Dresden, Germany}{jakob.piribauer@tu-dresden.de}{https://orcid.org/0000-0003-4829-0476}{}

\author{Robin Ziemek}{Technische Universität Dresden, Germany}{robin.ziemek@tu-dresden.de}{https://orcid.org/0000-0002-8490-1433}{}

\authorrunning{C. Baier et al.} 

\Copyright{Christel Baier, Clemens Dubslaff, Florian Funke, Simon Jantsch, Rupak Majumdar, Jakob Piribauer, and Robin Ziemek} 

\ccsdesc[100]{Theory of computation~Logic and verification}

\keywords{Model Checking, Causality, Responsibility, Counterfactuals, Shapley value} 

\relatedversion{} 

\acknowledgements{This work was funded by DFG grant 389792660 as part of TRR~248 -- CPEC (see \url{https://perspicuous-computing.science}), the Cluster of Excellence EXC 2050/1 (CeTI, project ID 390696704, as part of Germany's Excellence Strategy), DFG-projects BA-1679/11-1 and BA-1679/12-1, and the European Research Council under the
	Grant Agreement 610150 (\url{http://www.impact-erc.eu/}) (ERC Synergy Grant ImPACT).}

\nolinenumbers 

\EventEditors{Nikhil Bansal, Emanuela Merelli, and James Worrell}
\EventNoEds{3}
\EventLongTitle{48th International Colloquium on Automata, Languages, and Programming (ICALP 2021)}
\EventShortTitle{ICALP 2021}
\EventAcronym{ICALP}
\EventYear{2021}
\EventDate{July 12--16, 2021}
\EventLocation{Glasgow, Scotland (Virtual Conference)}
\EventLogo{}
\SeriesVolume{198}
\ArticleNo{1}

\usepackage[dvipsnames]{xcolor}
\usepackage{etex}
\usepackage{graphicx}
\usepackage{amssymb,amsmath,amsthm}
\usepackage[mathscr]{eucal}
\usepackage{verbatim}
\usepackage{comment}
\usepackage{mathtools}
\usepackage{amsfonts}
\usepackage{color}
\usepackage{graphicx}
\usepackage{float}
\usepackage{tikz}
\usepackage{xspace} 
\usepackage{pgf}
\usetikzlibrary{arrows,automata}
\usepackage[usenames,dvipsnames]{pstricks}
\usepackage{epsfig}
\usepackage{pst-grad} 
\usepackage{pst-plot} 
\usepackage{pstricks-add}
\usepackage[linesnumbered,ruled]{algorithm2e}
\usepackage{hyperref}
\usepackage[backgroundcolor=magenta!10!white]{todonotes}
\usepackage{enumitem}

\newcommand{\tudparagraph}[1]{\subparagraph*{#1.}}

\definecolor{darkgreen}{rgb}{0.1, 0.5, 0.1}

\def\reals{\mathbb{R}}

\def\set#1{{\{ #1 \}}}

\newcommand{\AS}{{\operatorname{AS}}}

\newcommand{\Sh}{{\operatorname{Sh}}}
\renewcommand{\Pr}{{\operatorname{Pr}}}

\newcommand{\playerone}{\texttt{SAT}}
\newcommand{\playertwo}{\texttt{UNSAT}}
\newcommand{\true}{\mathtt{true}}
\newcommand{\false}{\mathtt{false}}

\begin{document}

\maketitle

\begin{abstract}
	In view of the growing complexity of modern software architectures, formal models 
	are increasingly used to understand \emph{why} a system works the way it does,
	opposed to simply verifying \emph{that} it behaves as intended.
	This paper surveys approaches to formally explicate the observable behavior of reactive systems. 
	We describe how Halpern and Pearl's notion of actual causation inspired verification-oriented 
	studies of cause-effect relationships in the evolution of a system. 
	A second focus lies on applications of the Shapley value to responsibility ascriptions, 
	aimed to measure the influence of an event on an observable effect.
	Finally, formal approaches to probabilistic causation are collected and connected, 
	and their relevance to the understanding of probabilistic systems is discussed.
\end{abstract}

\section{Introduction}

Modern software systems are increasingly complex and even small changes to a system or its 
environment may lead to unforeseen and disastrous behaviors.
As software controls more aspects of our lives everyday, it 
is desirable---and for widespread acceptance in societal decisions, eventually inevitable---to 
have comprehensive and powerful techniques available to understand what a system does.

The field of formal methods has developed a portfolio of tools that provide confidence in the working of complex software systems.
In formal methods, one builds a formal model of a system and specifies its
desired behavior in an appropriate (temporal) logical formalism.
Algorithmic techniques such as model checking \cite{BK08,handbookMC} can answer 
the question whether the model satisfies the specification, or in other words, whether 
the system behaves as intended, often in a ``push button'' way. 
Moreover, an important aspect of these algorithms is that they can produce independently 
verifiable justifications of their outcome, such as \emph{counterexamples} or \emph{certificates}
to justify the violation or correctness of a property, respectively.
Since the earliest successes of model checking, the availability of counterexample traces was stated as a major
advantage for the method over deductive verification \cite{ClarkeGMZ95}.
As model checkers became more complex, concerns about their correct implementation led to research on producing
certificates for correctness. Examples are inductive invariants or derivations in a deductive system 
\cite{HenzingerJMNSW02,HofmannNR16,FunkeJB20,JantschFB20} that can be checked independently
from the verification process.

While certificates and counterexample traces can provide a useful explication about the behavior of a system,
they only provide rudimentary understanding of \emph{why} a system works the way it does.
In epistemic terms, the outcome of model checking applied to a system and a specification provides 
\emph{knowledge that} a system satisfies a specification or not in terms of
an assertion (whether the system satisfies the specification) and 
a justification (certificate or counterexample) to increase the belief in the result.
However, model checking usually does not provide \emph{understanding} on \emph{why} a system behaves in a certain way. 
Such an understanding can be obtained by causal links between possible events and their observed outcome.\footnote{Epistemologists since Gettier \cite{Gettier}
	will recognize that justified true belief does not constitute a comprehensive theory of knowledge.
	For the same reason, a theory of understanding as knowledge of causes is a matter of vigorous debate, 
	with Gettier-like counterexamples \cite{Grimm,PritchardKU,Pritchard}. 
	These subtle epistemic issues are orthogonal to our work.
}

The need to better understand why a system is correct or incorrect has led to a broad research program 
on models and reasoning methods that aim to provide such knowledge of causes~(see, e.g., 
\cite{Pearl09,Peters2017}).
The goal of this survey is to summarize research on causal reasoning in the field of verification
and highlight the challenges that lie ahead. 

The first step in understanding knowledge of causes is the mathematical formulation and study of \emph{causality}.
Grasping the intuitive concept of cause-effect relationships in a formal model has proved notoriously difficult. 
Centuries of philosophical reasoning on the subject have distilled the 
\emph{counterfactuality principle} \cite{Hume1739,Hume1748, Lewis1973} as a central feature of 
what constitutes an actual cause: 
if the cause had not occurred, then the effect would not have happened. 
While the counterfactuality principle was generally agreed upon, a rigorous mathematical formulation
was developed only recently, through the seminal work of Halpern and Pearl and their coworkers \cite{Pearl95,GallesP97,GallesP98,Halpern2000}.
In a nutshell, they model causal systems using \emph{structural equation models}, and provide a set of axioms to characterize
when an event is an actual cause of another.
We provide a summary of the foundations of causality and some of their applications in verification  in \Cref{sec:causality}.

While causality is a qualitative concept, in that an event is an actual cause of another or is not,
more recent work considers \emph{quantitative} measures of \emph{responsibility}.
Responsibility measures the relative importance that an event had in causing another event. 
In other words, the responsibility of an acting agent 
gauges what fraction of an observable effect can be attributed to that agent's behavior. 
Here, an agent could be, e.g., an individual, a coalition, a software component, or device in
a computer network.
Chockler and Halpern \cite{ChocklerH04} define the \emph{degree of responsibility} 
of an actual cause in the Halpern-Pearl sense  
based on the cardinality of the 
smallest witness change that makes an event a cause of another. 
A more recent strand for formalizing responsibility is based on the 
\emph{Shapley value} \cite{Shapley1953}. 
In cooperative game theory, the Shapley value measures the influence of an agent 
on the outcome jointly brought about by the agents and is classically used to find a fair 
division of a cost or a surplus among them. The appeal of the Shapley value stems on the 
one hand from its uniqueness with respect to a relatively simple set of axioms, and 
on the other hand from its seemingly universal applicability. Research on employing 
Shapley-like values for the explication of machine-learning predictions \cite{StrumbeljK14,DattaSZ16, LundbergL17,AasJL20,SundararajanN20} and the behavior 
of formal models \cite{YazdanpanahDJAL19, BaierFM2021, MascleBFJK2021, BaierFM2021b} is currently very active. 
A summary of applications of Shapley values in the verification context is provided in \Cref{sec:shapley}.

From a systematic viewpoint, causality and responsibility can be understood in either a \emph{backward} or a \emph{forward} manner \cite{vandePoel11}.
In the backward or \emph{ex post} setting, an effect has already transpired, and the goal is to describe 
its causes and determine their relative influence in producing the effect. 
The actual causation framework by Halpern and Pearl follows this paradigm, and therefore also most approaches presented in \Cref{sec:causality}. 
In the forward or \emph{ex ante} setting, a reasoning model includes possible contingencies 
and the goal is to characterize the global power of agents and events in affecting the outcome.
The forward responsibility in game structures (see \Cref{sub:games}) and the importance value for temporal logics (see \Cref{sub:importance}) pursue this pattern.
There are also attempts to express forward-looking causality notions by structural equations in the context of \emph{accountability} \cite{KaciankaIP20}. 
Seen from an operational angle, the distinction between backward and forward notions loosely relates to when the causal analysis is executed. 
Backward notions tend to be applicable at inspection time, e.g., to guide the debugging process in post mortem analyses. 
Forward notions are prone to be used at design time of the system, laying out general phenomena of its inner workings.

Finally, we consider causality in the setting of probabilistic models.
Unlike the deterministic setting, mathematical notions of causality and responsibility are less understood. 
There is widespread agreement in the philosophy literature that a quintessential characteristic of causes in the 
probabilistic setting is the \emph{probability-raising property} \cite{Suppes70, Cartwright79, Skyrms81, Eells91}: 
an occurrence of the cause should increase the probability of subsequently observing the effect. Nevertheless, it has also been observed that simply taking probabilities often leads to counterintuitive phenomena owing to mutual dependencies and latent correlations with other events \cite{Reichenbach1956,Cartwright79, Skyrms81}.
\Cref{sec:prob}  discusses attempts to formulate probability-raising approaches to causation for operational probabilistic models. 
These approaches tend to produce forward notions since probabilities inherently refer to a collection of evolutions in which an event happened.
There are numerous philosophical accounts on actual probabilistic causation \cite{Lewis1986,Kvart2004, Fenton-Glynn2017}. 
In terms of formal models, Pearl's early notion of actual causality in terms of \emph{causal beams} \cite{Pearl98, Pearl09} 
entailed probabilistic flavor, and the \emph{causal probabilistic logic} of \cite{VennekensDB09,BeckersV16} describes a language for reasoning about probabilistic causation. 
Nonetheless, we are aware of only a few works that study a probabilistic version of causality in operational models
(see, e.g., \cite{KleinbergM2009,abraham2018hyperpctl,DimFinkbeinerTorfah-probHyper-MDP-ATVA2020,BaierFJPZ21}). 
Along these lines, we point out open directions for research that focus on the operational point of view.

\section{Counterfactual Notions of Causality}
\label{sec:causality}

An important starting point for the study of causality in formal methods is the influential work by Halpern and Pearl 
\cite{HalpernP2001,HalpernP04,HalpernP05,Halpern15,HalpernBook}
on \emph{actual causality}, henceforth abbreviated \emph{HP causality}.
We provide a brief and informal overview of their definition.

Halpern and Pearl use \emph{structural equation systems} as a modeling language for causal models.
A causal model relies on 
exogeneous variables $U$ and endogeneous variables $V$, representing external or independent factors and 
internal factors, respectively. 
The value of each endogeneous variable $x\in V$ is specified by a deterministic function $f_x$ that 
may depend on exogeneous variables and on endogeneous variables that are preceding $x$ with respect to a
fixed order on $V$.
Intuitively, a causal model can be thought of as an arithmetic circuit whose primary inputs are the exogeneous
variables and where some of whose internal nodes are labeled by the endogeneous variables.
The circuit then specifies the functions defining the endogeneous variables as well as the 
dependencies between the variables.

More formally, let $M = (U, V, \set{f_x}_{x\in V})$ be a causal model. 
Given a formula $\varphi$ over the exogeneous and endogeneous variables (in some appropriate logic), 
and a \emph{context} $\vec{u}$ that assigns values to all variables in $U$, the goal of
actual causality is to state whether an assignment of values $\vec{X} = \vec{x}$ to a subset $X \subseteq V$ is a cause of $\varphi$.
Halpern and Pearl define $\vec{X} = \vec{x}$ to be a \emph{cause} of a formula $\varphi$ in 
$(M, \vec{u})$ if the following three axioms hold.
\begin{enumerate}[label=AC\arabic*:,leftmargin=*]
	\item both the cause and the effect are true: the model $(M, \vec{u})$ satisfies $\vec{X} = \vec{x}$ as well as $\varphi$,
	\item the principle of counterfactual dependence (discussed below), and
	\item causes are minimal: no partial assignment of $\vec{X}=\vec{x}$ satisfies AC1 and AC2. 
\end{enumerate}

The key to AC2 is captured by the notion of \emph{interventions}, describing a direct 
assignment of values to some endogeneous variables while disregarding their defining functions.
Formally, $[\vec{Y} \leftarrow \vec{y}]$ stands for the intervention on variables $Y \subseteq V$
by assigning them values $\vec{y}$ and leaving all other values for variables $V\setminus Y$
to follow from their defining functions. Then, $[\vec{Y} \leftarrow \vec{y}]\psi$ describes the impact
of an intervention on a formula $\psi$. An intervention thus can represent a counterfactual: 
\emph{what if} variables in $Y$ took values $\vec{y}$ instead of their actual values? 
Turning back to the definition of actual causes for $\varphi$, 
axiom AC2 now requires the existence of an intervention $[\vec{X} \leftarrow \vec{x}']$
on the variables in $X$ such that the effect $\varphi$ is not observable, i.e.,
$[\vec{X} \leftarrow \vec{x}']\neg \varphi$ holds in $(M,\vec{u})$.
The precise definition of AC2 is, however, more involved and several variants
exist for AC2 to account for different settings and applications.\footnote{Halpern and Pearl's definition of causality underwent a considerable 
	amount of development over the past 20 years, primarily varying AC2. 
	One usually distinguishes the \emph{original} version \cite{HalpernP2001}, 
	the \emph{updated} version \cite{HalpernP04,HalpernP05}, 
	and the \emph{modified} version \cite{Halpern15,HalpernBook} (alongside variations by other authors \cite{Hall07,Hitchcock01}). 
}

\subsection{Instances of HP Causality in Verification}\label{sec:instancehp}

Principles of causality have been used, often implicitly, in formal verification for a long time.
An early example is \emph{program slicing} (see, e.g., \cite{HarHie01u}) where 
by following program dependencies one aims to identify approximations of an actual cause 
for reaching a program location. Causality is also a key concept in error localization, 
the problem of reducing a counterexample trace for ease of debugging
\cite{Zeller02,BallNR03,RenieresR03,Groce04,WYIG06,JoseM11,ZhangMVJ17}. 
A correspondence of causality in counterexample traces to finding minimal $\playertwo$
cores has been identified in \cite{BeerBCOT09}.
Early and influential work on causality in formal verification is 
exemplified by research on \emph{vacuity} and \emph{coverage}.
Vacuity \cite{BeerBER97,KupfermanV99,PurandareS02}
explicates whether a positive verification result originates from an unintended trivial behavior. 
Coverage \cite{HoskoteKHZ99,ChocklerKKV01,ChocklerKV01,ChocklerKV06} is dual to vacuity, 
and explicates whether certain parts of the system were not relevant for the successful result.  
While for determining vacuity one considers small changes to the specification and
checks whether these change the result, coverage is obtained by perturbations to the system
rather than the specification and is actually a particular instance of HP causality.

Temporal logics play a crucial role in the verification context to describe
properties of and requirements on the system. Common temporal logics are, e.g., 
\emph{computation tree logic} (CTL) \cite{ClaEmeSis1986} or \emph{linear
temporal logic} (LTL) \cite{Pnu77a}. In LTL, e.g., $\neg E\;\mathcal{U}\,C$ 
describes that an effect $E$ does not occur before a cause $C$ and $\Diamond E$ 
stands for the effect $E$ to eventually occur.

\tudparagraph{From Coverage to HP Causality}
Coverage itself is a concept with a manifold of incarnations and we focus here on the
formalization by \cite{ChocklerHK2008}, where the connection of coverage to HP causality 
has been addressed. The operational model is provided by a Kripke structure $K$, i.e.,
a finite directed graph over states labeled by atomic propositions.
Further, we are given an atomic proposition $q$
and a specification $\psi$ expressed in an appropriate (temporal) logic over the set
of atomic propositions such that $K$ satisfies $\psi$.
Then, a state $s$ of $K$ is said to be \emph{$q$-covered} if changing the truth 
value of $q$ in $s$ leads to a structure that does not satisfy $\psi$. 
Considering the hypothetical mutant system in which $q$ takes the opposite value in $s$ 
corresponds to a counterfactual notion from the causality literature. 
Yet, coverage only allows simple counterfactuals containing \emph{individual} changes to the system.
As pointed out in~\cite{ChocklerHK2008}, it is for this reason that coverage at times fails to 
express deeper dependencies involved in the satisfaction of $\psi$.

To define a \emph{cause} in this setting, one can consider the following simple causal model: 
for each state $s$, there is one endogeneous variable $v_s$, which specifies whether the value 
of $q$ in $s$ is swapped in contrast to the original structure $K$, or not.
One first refers to a context where all variables $v_s$ are set to $\false$
and then considers possible swap operations. 
From $K$ and $\psi$ one can derive a Boolean function $\varphi$ over the endogeneous
variables $V$ such that an instantiation $I\colon V \to \{\true,\false\}$ satisfies 
$\varphi$ if and only if swapping the truth value of $q$ exactly in states $s$ 
with $I(v_s) = \true$ leads to a structure satisfying $\psi$.
Now $s$ is a \emph{cause of $\psi$ with respect to $q$}~\cite[Definition 3.2]{ChocklerHK2008} 
if there exists a set of variables $Y$ such that $[Y \leftarrow \true]\varphi$ and 
$[Y \cup \{v_s\} \leftarrow \true] \neg \varphi$ hold.
In other words, $s$ is a cause if there exists a set of states $S'$ (corresponding to variables $Y$) 
such that swapping $q$ in $S'$ leads to a structure satisfying $\psi$\footnote{The updated version 
of HP causality \cite{HalpernP04,HalpernP05} would require this condition also for all subsets of $S'$.}, 
but swapping the value of $q$ in $S'$ and $s$ gives a structure falsifying $\psi$.
These two conditions postulate precisely the axiom AC2, which takes a simpler form than 
usual thanks to the lack of higher-order dependencies among the variables in this causal model. 
In the presented causal model, axiom AC1 holds by the assumption that $K$ 
satisfies $\psi$, and the minimality axiom AC3 is trivially fulfilled as only single states are 
considered as potential causes.

While this causal model is very simple, in particular it does not include 
any dependencies in between variables, the work in~\cite{ChocklerHK2008} shows 
that even such restricted models are useful.

\tudparagraph{Fault Localization}
The causality interpretation of coverage presented above takes a forward-looking perspective 
in that changes to the system are globally tested against the given specification. 
In \cite{BeerBCOT09}, a similar approach is applied to the backward-oriented setting 
of fault localization, i.e., the problem of pointing out those parts of a (finite) 
counterexample trace $\pi$ that are most relevant for violation of a given linear-time
specification $\varphi$. In this incarnation of HP causality, the endogeneous variables $V$ 
contain a variable $v_{(s,q)}$ for each pair consisting of a state $s$ and atomic 
proposition $q$ of the Kripke structure. These variables can take values $\{\true,\false\}$ 
and the interpretation is exactly as before, namely $v_{(s,q)} = \true$ means that the 
truth value of $q$ in state $s$ is changed in contrast to the initial context. 
Moreover, the axiom AC2 takes the same form as in the previous case. 
Specifically, it expresses that there is a set of variables $Y\subseteq V$ such that changing 
the truth value for the corresponding state-proposition pairs lets $\pi$ still 
violate $\varphi$, while additionally swapping $q$ in $s$ leads to $\pi$ satisfying 
$\varphi$ (interpreted over a weak semantics of LTL on finite paths).

\tudparagraph{Counterfactual Reasoning for Configurable Systems}
Nowadays, almost every practical software system is configurable, let it be using \#\texttt{ifdef}
constraints or through \emph{features}~\cite{Apel2009AnOO}. 
Features inherently have a designated meaning, usually expressed by their name, e.g., 
a ``verbose'' feature indicates that the software will expose additional information during runtime. 
Debugging configurable systems is challenging, as the number of possible systems
	suffers from an exponential blowup in the number of features.
	While there are specifically tailored methods for analyzing configurable
	systems~\cite{Thum14}, e.g., through model checking~\cite{Cla2010,DBK15},
	research on identifying root causes in configurable systems on the 
	abstraction level of features is still in its infancy. 
	Such a causal analysis can provide useful insights for debugging: developers can focus on
	the parts implementing the features identified to be responsible for the bug, and
	users can obtain suggestions to reconfigure the system to not expose the bug.
First ideas to explicate which feature activations and deactivations cause an effect 
in configurable systems were described in \cite{CeTIBook-2020-Chap13}.
There, the set of feature configurations with observable effect
is obtained by configurable systems analysis, e.g., through family-based 
verification~\cite{Cla2010,Thum14,DBK15,CDKB18}.
Exploiting the Boolean case of HP causality~\cite{EiterL2004,IbrahimP20}, 
those partial feature configurations can be determined where the corresponding systems 
all show the effect (see AC1), for which there is a reconfiguration
that does not exhibit the effect (AC2), and that are minimal (AC3). 

\subsection{Further Approaches Inspired by HP Causality}

The work \cite{DattaDKSS15} presents a formal definition of actual causes in the 
setting of concurrent interacting programs. Originating from logs written by the 
concurrent system, the goal is to localize causes in those \emph{program actions} 
that are most relevant for the violation of a desirable property. 
The approach is investigated in detail for the prominent class of safety properties, 
with a view towards legal accountability in security-critical systems.

In~\cite{KoeblLS20}, a causality-based approach to explain \emph{timed diagnostic traces} 
has been presented, which are used as counterexamples for model-checking results in timed systems.
Such traces represent a set of violating executions and the goal of~\cite{KoeblLS20} is 
to compute the parts that can be considered causal for violating the property.

A different definition inspired by HP causality was used in \cite{Leiner-FischerL2013}.
There, causes for reachability properties are formulas of a temporal logic called 
\emph{event order logic}, used to describe temporal relations between events.
Algorithms to compute causes in this sense were also studied in~\cite{BeerHKLL15,KoeblL19}, 
and the approach was extended to handle general LTL formulas as effects rather than just reachability
in~\cite{CaltaisGL2018}.

In~\cite{GoesslerL14,GoesslerL15} the authors argue that the HP causality, which is propositional 
in nature, is not the ideal starting point for a framework of causality in formal verification.
They present a formalism which is based on counterfactual reasoning, uses system 
traces as first-class objects and is designed to work for compositional systems. 
In~\cite{GoesslerS20} the formalism is further generalized by defining abstract 
\emph{counterfactual builders}, which specify what alternative scenarios should 
be considered for counterfactual reasoning.
Further,~\cite{GoesslerS20} also considers hyperproperties as specifications.
While hyperproperties are useful to specifying system properties, it was observed 
in~\cite{Coenen19} that they can also be used to formalize causality.
Similar observations have been made  for probabilistic causation~\cite{abraham2018hyperpctl,DimFinkbeinerTorfah-probHyper-MDP-ATVA2020}.


\section{Responsibility and Shapley-like Ascriptions}
\label{sec:shapley}

While the previous section defined and applied qualitative concepts of causation, 
this section shifts the focus towards quantitative approaches of \emph{responsibility}. 
Loosely speaking, responsibility refers to a numerical value designed to measure how much weight an event had in 
producing an effect, relative to concurring or competing events linked to the same effect. 
There is widespread agreement that a necessary condition for assuming responsibility is causal relevance of the event 
in question to the effect \cite{Feinberg1970, BrahamvanHees2012}. 
As a consequence, the term \emph{responsibility} usually builds directly or indirectly on concepts of causality.
While the specific numerical value in a notion of responsibility may not have a semantic content, 
it can order the events in terms of their causal relevance.

Chockler and Halpern \cite{ChocklerH04} introduced the notion of \emph{degree of responsibility},
which is attributed to actual causes in causal models of HP causality.
This degree measures how many changes to the evolution of events are necessary 
until counterfactual values for the actual cause change the observable effect. 
In \cite{ChocklerHK2008} this notion is combined with the study of mutant coverage to 
build a degree of responsibility in CTL model checking assigned to state-proposition 
pairs (see also Section~\ref{sec:instancehp}).

The degree of responsibility measures the influence of an event by looking at how many 
further counterfactual changes are (minimally) required to swap the effect, 
but is does not take into account how many such minimal sets of changes exist. 
One can argue that a cause is individually more influential if it admits \emph{many} 
such sets since this means less dependencies on other events. 
This rationale has generated an active strand in formalization of responsibility 
based on the \emph{Shapley value} \cite{Shapley1953}. The Shapley value is a central 
solution concept from theoretical economics and was originally designed to find a fair 
distribution of a financial surplus that was brought about cooperatively by a number of producers. 

Formally, a \emph{cooperative game} with $n$ players is a mapping $g\colon 2^{[n]}\to \reals$ 
such that $g(\emptyset) = 0$, where $[n] = \{1,\ldots, n\}$. The value $g(C)$ is meant to 
represent the surplus (or, depending on the specific situation, the cost) that the 
coalition $C\subseteq[n]$ can ensure upon acting collaboratively. 
The Shapley value of player $i$ is then defined as
\begin{equation}\label{eq:Shapley} 
\Sh(i) \enskip =\enskip \frac{1}{n!} \cdot\sum_{\pi \in S_n} g(\pi_{\geq i}) - g(\pi_{\geq i} \setminus \{i\})
\end{equation}
where $S_n$ denotes the set of self-bijections $[n]\to [n]$ and where 
$\pi_{\geq i} = \{j\in[n] \mid \pi(j)\geq \pi(i)\}$ for a given $\pi\in S_n$. 
Intuitively, $g(\pi_{\geq i}) - g(\pi_{\geq i} \setminus \{i\})$ describes the marginal 
contribution of player $i$ to the coalition $\pi_{\geq i}$. The Shapley value takes the 
average of all such marginal contributions. Thus, $\Sh(i)$ is a measure for
the overall influence of player $i$ in the game $g$.

The general setup of cooperative games as real-valued functions on the power set of 
$[n]$ makes the Shapley value amenable to measuring the influence of abstract players 
in formalized situations of collaborative interaction. This rationale has recently been 
invoked for the interpretation of machine learning models \cite{StrumbeljK14,DattaSZ16, LundbergL17,AasJL20,SundararajanN20}. 
In this case, the players are the input parameters to a machine learning model, and the Shapley 
value has the goal to measure the influence of each parameter on the output of the model. 

This section outlines three approaches that employ the Shapley value as a means to distribute 
an overall effect into individual responsibilities. In \Cref{sub:games} the general setting of 
an \emph{extensive form game} is chosen and responsibilities are attributed to its players 
with respect to producing a certain outcome. \Cref{sub:importance} discusses a notion of 
importance of states for the satisfaction of an LTL property in a Kripke structure. 
\Cref{sub:cont} finally presents an extension of the Shapley value that can be used to 
define responsibilities in a setting of continuously varying parameters.

\subsection{Responsibility in Game Structures}
\label{sub:games}

As summarized in \Cref{sec:causality}, causal models are by now a fundamental building 
block for notions of actual causation in the verification domain. 
However, in complex scenarios that involve cooperative interaction,  non-cooperative 
competition, and imperfect information, they fall short of modeling various natural 
features such as temporal sequentiality, knowledge, and agency. 
The work \cite{BaierFM2021b} presents an approach to establish notions of responsibility 
in these strategic settings by passing to \emph{extensive form games} \cite{vonNeumann1928, Kuhn}. 
These provide a popular formalism for studying the dynamics that underlie strategic 
interaction in the presence of competing objectives. In a nutshell, an extensive form 
game is an explicit presentation of the strategic scenario in terms of a tree structure 
whose edges describe the transitions between states when actions are taken, certain states 
may be indistinguishable for the players given their knowledge, and each path from the root 
to a leaf is associated with an outcome. Apart from being a highly expressive model, a 
century of research on the subject has generated a rich set of solution concepts 
on which a study of responsibility can build, primarily following economic rationales.

In \cite{BaierFM2021b} three responsibility notions are defined with respect to an event $E$ that is encoded by a binary labelling on the leaves of the game tree, i.e., $E$ took place on a play or not. 
All three notions follow the common two-step process consisting of first defining (qualitatively) what it means for a coalition $C$ to be responsible and then extracting (quantitatively) an individual responsibility value through an application of the Shapley value on coalition responsibilities. That is, one takes the cooperative game $g$ to take binary values $\{0,1\}$ depending on whether or not a coalition is responsible. 
They also share the counterfactual paradigm in that a necessary condition for being responsible for the occurrence of $E$ is the power to preclude $E$.
While the notions can be ordered according to their logical strength, they are perhaps best explained along two lines of distinction given by the \emph{temporal perspective} and \emph{epistemic state}.

The temporal perspective can be either \emph{forward-looking} or \emph{backward-looking} \cite{vandePoel11}. For forward-looking notions one attains a prospective, \emph{ex ante} viewpoint that studies the preclusive power for the game as a whole. The forward-looking notion put forth in \cite{BaierFM2021b} is called \emph{forward responsibility} and requires the coalition to possess a strategy that globally avoids $E$. In contrast, backward-looking notions consider a specific play from a retrospective, \emph{ex post} viewpoint and study who was responsible for $E$ as the play evolved. 

Depending on how the epistemic state is taken into account, \cite{BaierFM2021b} distinguishes \emph{strategic} backward responsibility and \emph{causal} backward responsibility. 
In order for a coalition to be strategically backward responsible, it must have had the power to avoid $E$ at some point on the play and it must have been aware of this fact \emph{given its epistemic knowledge}. 
In situations of imperfect information, this latter condition is crucial for arriving at a \emph{responsibility-as-capacity} notion \cite{vandePoel11} in a strategic sense that goes beyond a mere counterfactuality check: 
when one does not know all relevant information, one can even bring about $E$ inadvertently or unintendedly. 
Causal backward responsibility essentially drops the latter requirement in that the coalition is able to avoid $E$ from some point on, 
everything else held fixed. This corresponds to the \emph{responsibility-as-cause} from the 
classification presented in \cite{vandePoel11}.

There is a translation of a causal model into an extensive form game under which causal backward 
responsibility corresponds exactly to but-for causes \cite{BaierFM2021b}. It can therefore happen 
that a player is causally backward responsible without belonging to an actual cause in the HP 
causality sense \cite{HalpernP04} and, therefore, with degree of responsibility $0$ in the 
sense of \cite{ChocklerH04}. In the prototypical example in which Suzy and Billy both 
throw rocks at a bottle and Suzy's stone hits first, Billy's degree of responsibility is $0$, 
while both are attributed causal backward responsibility $1/2$. Since both players acted in 
exactly the same way based on the same information, there is reason to  favor the latter 
symmetric notion, and avoid actual causation \emph{en route} to accurately model 
intuitive responsibility concepts.
A detailed comparison to causal models, other notions of responsibility in strategic 
games of imperfect information \cite{YazdanpanahDJAL19}, and proof-theoretic 
approaches to formalize responsibility \cite{Broersen11,NaumovT20} is given in \cite{BaierFM2021b}.

\subsection{The Importance Value for Temporal Logics}
\label{sub:importance}

The paradigm of passing from binary coalitional responsibilities to quantitative individual 
responsibilities by virtue of the Shapley value is also applied in \cite{MascleBFJK2021} to model check
Kripke structures against temporal logic specifications. 
The resulting notion is called the \emph{importance value} and measures the influence of a state in a system for the satisfaction of a given specification.
Intuitively, a state is important in this framework if the way that the nondeterministic choices of the state are resolved has a large impact on whether the given specification is met.

Formally, let $K$ be a Kripke structure with states $S$ and a dedicated initial state, and 
$\varphi$ be an LTL formula. Then one defines the cooperative game $g\colon 2^S\to \{0,1\}$ using an induced two-player game as follows.
For a set of states $C \subseteq S$ we let ${\cal G}_C$ be the two-player game over the arena $K$ 
where player \playerone{} controls the states in $C$, player \playertwo{} controls the states in 
$S {\setminus} C$ and the winning condition is $\varphi$.
Then, $g(C) = 1$ if player \playerone{} wins $\cal{G}_C$, and $g(C) = 0$ otherwise. 
With this definition, the importance value $\mathcal{I}(s)$ of a state $s \in S$ with respect to $K$ and $\varphi$ is defined to be the Shapley value of player $s$ in $g$ (see \Cref{eq:Shapley}). The notion can be straightforwardly extended to define the importance of a set of states $P_i \subseteq S$, where $S = P_1 \:\dot\cup\:\ldots\: \dot\cup\: P_n$ is a given partition of the state space. This generalization is intended to take an existing compositional structure of the system appropriately into account.

The work \cite{MascleBFJK2021} studies the associated computational problems of deciding whether $\mathcal{I}(s) > 0$ (called the \emph{usefulness problem}) and deciding whether  $\mathcal{I}(s) > \eta$ for a rational threshold $\eta$.
The intrinsic complexity of solving two-player LTL-games (the decision problem is 2EXPTIME-complete) carries over to these problems. 
This computational intractability of the importance value motivates further studying the complexity 
when restricted to fragments of LTL, and tight complexity results were shown in \cite{MascleBFJK2021}
for a wide range of specifications.

In \cite{MascleBFJK2021}, the presented framework is also applied to CTL model checking of 
\emph{modal transition systems} (MTS) \cite{LarsenT1988a}.
MTSs have two levels of nondeterminism: the standard nondeterminism of the underlying graph
and additionally a choice on which of the transitions in a state are actually included in the system.
The latter kind of nondeterminism is used to design a two-player game where one player 
tries to satisfy the CTL specification and the other player tries to violate it.
However, since the semantics of CTL relies on infinite trees and the order in which the 
branches are evolving has a strong impact on which player wins, there does not appear to 
be a natural candidate game that proceeds in a turn-by-turn fashion. 
Hence~\cite{MascleBFJK2021} considers \emph{one-shot games} in which the players commit to a 
valid set of transitions in the states under their control once at the beginning of their play.
This determines once more a binary cooperative game $g$ that induces importance values in the same way as for LTL.

There is a straightforward generalization of the importance value 
to a $2\frac{1}{2}$-player game $\mathcal{G}$
in which the actions taken by the players are associated with probability distributions over the states. 
In this formalism, the players each make non-deterministic choices among its available actions, 
but the actual successor state then depends on a random choice according to the associated distribution. 
Given an LTL specification, the goal of \playerone{} is to maximize the probability 
that the resulting path satisfies the specification, while \playertwo{} tries to minimize it. 
These $2\frac{1}{2}$-player games are \emph{determined} in a quantitative sense \cite{Martin98}: 
the maximal probability that can be enforced by \playerone{} against all strategies of \playertwo{} 
is $1$ minus the minimal probability that can be enforced by \playertwo{} against all 
strategies of \playerone. This probability is called the \emph{value} $\operatorname{val}(\mathcal G)$ 
of the game (see also the survey \cite{ChatterjeeH12}).
Let $S =S_{\playerone}\:\dot\cup\: S_{\playertwo}$ be the partition of the states of $\mathcal G$ 
into those under control of \playerone{} and \playertwo, respectively. For a subset 
$C\subseteq S_{\playerone}$ the value $g(C)$ is then defined as $\operatorname{val}(\mathcal{G}_C)$, 
where $\mathcal{G}_C$ is the $2\frac{1}{2}$-player game obtained from $\mathcal{G}$ by putting 
the states in $S {\setminus} C$ under the control of player \playertwo. 
Taking Shapley values as above then 
induces the importance value of a state in $S_{\playerone}$.

\subsection{Attributing Responsibility in Continuous Models}
\label{sub:cont}

The Shapley value \cite{Shapley1953} is an inherently discrete solution concept. 
On the other hand, realistic formal models of reactive systems often entail continuous features such as timing \cite{AlurD1994,Dill1990,Lewis90}, physical phenomena \cite{AlurCHHHOS95, Sproston00}, or parametric dependencies \cite{JonssonL1991, GivanLD2000, KozineU2002}. 
Notions of responsibility for these models therefore tend to require new mathematical approaches if the continuous nature is to be taken into account appropriately. 

The continuous scenario seen from an economic angle generalizes (discrete) cooperative games: 
rather than just participating in a coalition, the $n$ players of a game each pick a 
value $v_i$ from a continuous domain $D_i\subseteq\mathbb{R}$ including $0$ and the (generalized) game then determines a collective surplus or cost based on this input. 
This is formally described by a continuously differentiable function $g\colon D_1\times \ldots \times D_n\to \mathbb{R}$ such that $g(0,\ldots, 0) = 0$. 
Economists usually take the domains to be of the form $D_i = [0, m_i)$ for some maximal input $m_i\in \mathbb{R}$, and further assume $g$ to be non-decreasing with non-negative range. 
The \emph{Aumann-Shapley value} \cite{AumannS1974} is a generalization of the Shapley value designed to provide a solution to the question how the value $g(v_1,\ldots, v_n)$ should be ``fairly'' distributed among the players. It is one instance of what is called a \emph{cost-sharing scheme} and admits an axiomatization in the spirit of its discrete predecessor \cite{FriedmanMoulin, SundararajanN20}.

Inspired by this model, the work \cite{BaierFM2021} presents an approach to measure 
the relative importance of the parameters on the behavior of \emph{parametric Markov chains}
for a wide range of properties, including $\omega$-regular specifications, specifications
in probabilistic CTL, and on expected rewards.
Here, a parametric Markov chain is a directed graph where each edge is assigned a probability that may depend on a set of parameters  such that for each instantiation of the parameters
the probabilities outgoing from a state sum up to $1$. A parametric
Markov chain instantiated with fixed values for the parameters then coincides with a \emph{discrete-time Markov chain} (DTMC).
For this purpose the aforementioned assumptions on $g$ must be relaxed: the continuously 
differentiable function $g\colon D\to \mathbb{R}$ has arbitrary domain $D\subseteq\mathbb{R}^n$ 
and is not subject to monotonicity and non-negativity restrictions.  This also means that the 
canonical baseline value $0$ for $v_i$ is not always available anymore. The responsibility problem 
in this generalized setting then reads as follows: given $g$ and two parameter choices $v, v'\in D$, 
how \emph{responsible} is the $i$-th parameter for the observable change $g(v') - g(v)$? 
In this slightly generalized form, the Aumann-Shapley value of the $i$-th parameter is defined as
\begin{equation}\label{eq:AS}
\AS_i(g, v, v') = (v'_i-v_i)  \cdot\int_0^1 \partial_i g\, (v + \alpha (v' - v))\:d\alpha.
\end{equation}
The integrand involves the $i$-th partial derivative of $g$ and intuitively measures the 
marginal contribution of the $i$-th parameter at the points lying between $v$ and $v'$. 
The integral then takes the average of these contributions along the straight line from $v$ to $v'$.
While taking the straight line is desirable in an economic context to meet the 
\emph{average cost for homogeneous goods axiom} \cite{FriedmanMoulin}, this axiom is 
often void of meaning when applied to formal systems. When one replaces the straight line 
in \Cref{eq:AS} with an arbitrary (monotonic) path from $v$ to $v'$, then one speaks of 
\emph{path attribution schemes} \cite{Sun_2011}. Of course, taking different paths 
induces different attributions, and which ones should be considered worthwhile 
depends on the specific scenario. This could for instance be due to potential 
restrictions on the way that changes on the parameters can be implemented in practice. 
The work \cite{BaierFM2021} applies these path attribution schemes to the function induced 
by $\omega$-regular or probabilistic CTL specifications on a parametric Markov chain. 
The set of axioms presented there is adjusted to this particular situation and justifies why one can conceive the value $\AS_i(g, v, v')$ as the fraction of the observable effect $g(v') - g(v)$ that is produced by the $i$-th parameter.

It is noteworthy, however, that the approach put forth in \cite{BaierFM2021} is by no means 
specific to the context of parametric Markov chains. Any scenario in which continuously varying parameters 
determine a value can in principle be handled similarly. Of course, which path attribution 
schemes should be regarded as meaningful needs to be checked case-by-case, and corresponding 
axiomatizations should be chosen with care. 
But it is no accident that the main decidability result in \cite{BaierFM2021} is formulated in terms of path 
attribution schemes on functions in $n$ independent variables---a generality that makes the approach potentially applicable for a range of similar problems.

\section{Probabilistic Causation}
\label{sec:prob}

As seen in the preceding sections, notions of causality and responsibility have been widely explored  in the non-probabilistic setting.
In contrast, there have been far less attempts at defining a suitable notion of causes for probabilistic operational systems
such as Markov chains.
However, probabilistic theories of causation have been considered in various philosophical accounts \cite{Suppes70, Cartwright79, Skyrms81, Eells91}.
One central idea behind these theories is the \emph{probability-raising principle}, which goes back to Reichenbach \cite{Reichenbach1925, Reichenbach1956}.
It states that causes should raise the probability of their effects.
After observing a cause $C$, the probability of an effect $E$ is higher than after observing that the cause has not occurred. Formulated with conditional probabilities, this can be written as
\[
\Pr(E\mid C) > \Pr(E \mid \neg C),\qquad\text{ or equivalently }\qquad\Pr(E\mid C)>\Pr(E).
\]
For the conditional probabilities to be well-defined, it is necessary that $\Pr(C)>0$ and $\Pr(\neg C)>0$. Later on, we will make sure that the events conditioned on have positive probability. 
Note that if $\Pr(C)>0$ and $\Pr(E\mid C)>\Pr(E)$, it already follows that  $\Pr(\neg C)>0$. 
Defining $p\stackrel{\text{\tiny def}}{=} \Pr(C)$, the equivalence of the two inequalities follows from the equation $\Pr(E)=p\cdot \Pr(E\mid C) + (1-p) \cdot \Pr(E\mid \neg C)$,
which implies \[ \Pr(E\mid C)- \Pr(E) = (1-p)  (\Pr(E\mid C)-  \Pr(E\mid \neg C)).\]

The probability-raising principle alone, however, cannot distinguish between cause and effect as it holds if and only if $\Pr(C\mid E)>\Pr(C\mid \neg E)$ as well.
For this reason, additional conditions have to be imposed for causal reasoning.
One key condition is temporal priority, which prescribes that a cause has to occur \emph{before} the effect.

This section formalizes both the probability-raising principle as well as the requirement of temporal priority for probabilistic operational models. 
We draw connections between different ideas from the literature to provide an overview over basic probabilistic notions of causality in the context of formal verification. 
For this, we assume to have given a DTMC $\mathcal{M}$ with a probability distribution over initial states. 
This way, the sets $\Pi_\varphi$ of paths starting in initial states and fulfilling an LTL property $\varphi$
are measurable~\cite{Vardi85} and have a well-defined probability value $\Pr_{\mathcal{M}}(\Pi_\varphi)$,
which we also denote by $\Pr_{\mathcal{M}}(\varphi)$.
Applying the probability-raising principle and expressing the temporal priority 
using LTL leads to the following first definition of causality in DTMCs
for reachability properties.

\begin{definition}[reachability-cause]
\label{def:cause_Reichenbach_reachability}
	Let $\mathcal{M}$ be a DTMC with state space $S$ and let $C,E\subseteq S$ be two disjoint sets of states.
	Then $C$ is a \emph{reachability-cause} of $E$ if $\Pr_{\mathcal{M}}(\neg E \, \mathcal{U} \, C)>0$ and
	\begin{equation}\label{eq:reachCause}
		\Pr_{\mathcal{M}}(\Diamond E \mid \neg E \; \mathcal{U}\, C) > \Pr_{\mathcal{M}}(\Diamond  E).
	\end{equation}
\end{definition}

Note that  \Cref{eq:reachCause} implies that $\Pr_\mathcal{M}\big( \neg (\neg E \; \mathcal{U}\, C)\big)>0$. This ensures that also $\Pr_{\mathcal{M}}\big(\Diamond  E \mid  \neg (\neg E \; \mathcal{U}\, C)\big)$ is well-defined and so
 \Cref{eq:reachCause}
 is equivalent to $\Pr_{\mathcal{M}}(\Diamond E \mid \neg E \; \mathcal{U}\, C) > 
\Pr_{\mathcal{M}}\big(\Diamond  E \mid  \neg (\neg E \; \mathcal{U}\, C)\big)$.
If there are no paths first reaching the effect $E$ and afterwards the cause $C$,
 e.g., because the states in the effect $E$ are absorbing,
  \Cref{eq:reachCause} simplifies to 
$
	\Pr_{\mathcal{M}}(\Diamond E \mid \Diamond C) > \Pr_{\mathcal{M}}(\Diamond  E )
	$.
	
In this treatment of reachability properties, a cause $C$ specifies the set of finite executions ending in $C$ that cause the subsequent extension to an infinite execution to satisfy $\Diamond E$.
This idea can be lifted to the treatment of causes of arbitrary events in $\mathcal{M}$ specified by a measurable set of infinite paths $\mathcal{L}\subseteq S^\omega$.
A cause is then a set of finite paths $\Gamma\subseteq S^+$. 
Besides the probability-raising property, the temporal priority condition needs to be included. For path properties this needs extra consideration.
While for a cause $\Gamma\subseteq S^+$ it is clear that the cause is observed once a finite path in $\Gamma$ is generated in a DTMC, this is not the case for the effect $\mathcal{L}\subseteq S^\omega$ as it consists of infinite executions.
However, it seems natural to say that the effect occurred on a finite path $\delta$ whenever $\Pr_{\mathcal{M}}(\mathcal{L}\mid \delta)=1$, i.e., if a generated finite path ensures that almost all infinite extensions belong to $\mathcal{L}$.
Here, we used $\delta$ to also denote the event of all infinite paths having $\delta$ as a prefix.
Analogously, for a set of finite paths $\Gamma$, we denote by $\Gamma$ also the event of all infinite paths with a prefix in $\Gamma$.
Consequently, $\neg \Gamma$ denotes the event of all infinite paths that have no prefix in $\Gamma$.
The discussed treatment of temporal priority is now used in the following definition of a cause in a DTMC.

\begin{definition}[global PR-cause]
	\label{def: Reichenbach causality}
	Let $\mathcal{M}$ be a DTMC with state space $S$, let $\Gamma\subseteq S^+$ be a non-empty 
	set of finite paths, and let $\mathcal{L}\subseteq S^\omega$ be a measurable set of paths.
	Then, $\Gamma$ is a \emph{global probability-raising cause (global PR-cause)} for $\mathcal{L}$ if the following
	two conditions hold:
	\begin{description}[style=multiline,leftmargin=1.7cm,font=\normalfont\color{black}]
		\item[PAC1:] $\Pr_{\mathcal{M}} (\mathcal{L} \mid \Gamma) > \Pr_{\mathcal{M}} (\mathcal{L}) $, and
		\item[PAC2:] for all $\gamma\in \Gamma$, no proper prefix $\gamma^\prime$ of  $\gamma$ satisfies $\Pr_{\mathcal{M}}(\mathcal{L}\mid \gamma^\prime)=1$.
	\end{description}
\end{definition}
As $\Gamma$ is a set of finite paths in $\mathcal{M}$, the cylinder set spanned by each $\gamma^\prime \in \Gamma$ has positive probability. 
So, the conditional probabilities
$\Pr_{\mathcal{M}}(\mathcal{L}\ | \ \Gamma)$ and
$\Pr_{\mathcal{M}}(\mathcal{L}\ | \ \gamma^\prime)$ 
  in Definition \ref{def: Reichenbach causality} are well-defined.
  
  Axiom PAC1 expresses the 
  probability-raising principle. It implies that $\Pr_{\mathcal{M}}(\neg \Gamma)>0$. As this ensures that all necessary conditional probabilities are well-defined, PAC1 is equivalent to 
  the probability-raising condition
$\Pr_{\mathcal{M}} (\mathcal{L} \mid \Gamma) > \Pr_{\mathcal{M}} (\mathcal{L}\mid \neg \Gamma) $. 
Axiom PAC2 captures that the effect must not occur before the cause.

The requirement on the cause in the provided definition is of \emph{global} nature: 
the cause $\Gamma$ as a whole has to guarantee the probability-raising property with respect to the effect.
Single elements $\gamma\in \Gamma$, however, do not necessarily guarantee that the probability of the effect has been raised.
Furthermore,
under mild assumptions, the definition subsumes the treatment of reachability properties above:
\begin{proposition}
	Let $\mathcal{M}$ be a DTMC with state space $S$ and let $C,E\subseteq S$ be disjoint. 
	Assume that no state in $s\in S\setminus E$ satisfies $\Pr_{\mathcal{M},s}(\Diamond E)=1$. 
	Then the following are equivalent:	
	\begin{enumerate}
		\item $ C$ is a reachability-cause for $ E$. 
		\item The set $\Gamma$ of finite paths in $(S\setminus (C\cup E))^\ast C$ is a global PR-cause for the set $\mathcal{L}$ of  paths satisfying $\Diamond E$.
	\end{enumerate}
\end{proposition}
In the proposition above, $\Pr_{\mathcal{M},s}$ denotes the probability measure induced by $\mathcal{M}$ with assuming 
$s$ as the unique initial state.
A related notion of causality based on probability-raising in DTMCs has been introduced by Kleinberg et al. in a series of papers \cite{KleinbergM2009, KleinbergM2010,Kleinberg2011,HuangK15,ZhengK2017}.
Here, probabilistic CTL is used to describe the cause $C$ and the effect $E$ via state formulas. 
We can describe both events also directly as sets of states in the DTMC by considering exactly 
those states that fulfil the corresponding probabilistic CTL formula. 
For reachability properties, the set  $C$ is then said to be a cause of $E$ if
$\Pr_{\mathcal{M},c}(\Diamond E) > \Pr_{\mathcal{M}}(\Diamond E)$ for all $c\in C$. 
So, the requirement for this notion of causality is \emph{local}: reaching any state $c\in C$ has to ensure that the probability of reaching $ E$ afterwards is raised.
In case $C=\{c\}$ is a singleton disjoint from $E$, this notion agrees with Definition \ref{def:cause_Reichenbach_reachability}.

Adapting PAC1 to sets of paths and including the temporal priority requirement that the effect 
does not occur before the cause (PAC2 as before), we obtain the following definition of causality:
\begin{definition}[local PR-cause] \label{def: Kleinberg Causality rect}
	Let $\mathcal{M}$ be a DTMC with state space $S$, $\Gamma\subseteq S^+$ a set of finite paths, and let $\mathcal{L}\subseteq S^\omega$ be a measurable set of paths.
	Then $\Gamma$ is a \emph{local probability-raising cause} (local PR-cause) for $\mathcal{L}$ if
	\begin{description}[style=multiline,leftmargin=1.7cm,font=\normalfont\color{black}]
		\item[PAC1\textsuperscript{loc}:] for all $\gamma\in \Gamma$ we have $ \Pr_{\mathcal{M}} (\mathcal{L} \mid \gamma) > \Pr_{\mathcal{M}} (\mathcal{L})$, and 
		\item[PAC2:] for all $\gamma\in \Gamma$ no proper prefix $\gamma^\prime$ of  $\gamma$ satisfies $\Pr_{\mathcal{M}}(\mathcal{L}\mid \gamma^\prime)=1$.
	\end{description}
\end{definition}

Axiom PAC1\textsuperscript{loc} can be seen as the local version of PAC1. Clearly, PAC1\textsuperscript{loc} implies PAC1.
Furthermore, PAC1\textsuperscript{loc} implies that $\Pr_{\mathcal{M}}(\neg \gamma)>0$ for all $\gamma \in \Gamma$. 
Hence, we could equivalently reformulate PAC1\textsuperscript{loc}  as
$ \Pr_{\mathcal{M}} (\mathcal{L} \mid \gamma) > \Pr_{\mathcal{M}} (\mathcal{L}\mid \neg \gamma)$ for all $\gamma\in \Gamma$.

The work by Kleinberg et al. proceeds relative to an explicit probability value $p>\Pr_{\mathcal{M}}(\mathcal{L})$ such that 
$\Pr_{\mathcal{M}}(\mathcal{L}\mid \gamma)\geqslant p$ for all $\gamma$ in a cause $\Gamma$. The higher this value $p$ lies above $\Pr_{\mathcal{M}}(\mathcal{L})$, 
the greater is the amount by which all elements of $\Gamma$ are guaranteed to raise the probability of the effect $\mathcal{L}$.

Such a reference to a specific threshold value $p$ has also been incorporated into a notion of \emph{$p$-causes} 
in \cite{BaierFJPZ21}. Motivated by monitoring applications~(see, e.g., \cite{LeuSch09g}),
the underlying idea is that notions of causality could be used to foresee undesirable behavior. 
If a cause for an erroneous execution is observed, countermeasures can be taken before the error actually occurs. 
Here it is particularly useful to specify a sensitivity $p$ that expresses how likely an error is after observing the cause.
In addition, the occurrence of an erroneous execution should not stay undetected. 
Therefore, an additional condition is imposed on $p$-causes: 
almost all executions that exhibit the error should have a prefix in the cause.
Together with the temporal priority of the cause (PAC2) as before, these requirements lead to the following definition:

\begin{definition}[$p$-cause]
	Let $\mathcal{M}$ be a DTMC with state space $S$ and $p\in(0,1]$. A non-empty set $\Gamma\subseteq S^+$ is a \emph{$p$-cause} for a measurable set $\mathcal{L}\subseteq S^\omega$ if
	\begin{description}[style=multiline,leftmargin=1.7cm,font=\normalfont\color{black}]
		\item[PAC1\textsuperscript{p}:] for all $\gamma\in \Gamma$ we have $\Pr_{\mathcal{M},s_0}(\mathcal{L}\mid \gamma)\geqslant p$, 
		\item[PAC2:] for all $\gamma\in \Gamma$ no proper prefix $\gamma^\prime$ of  $\gamma$ satisfies $\Pr_{\mathcal{M}}(\mathcal{L}\mid \gamma^\prime)=1$, and
		\item[PAC3:] $\Pr_{\mathcal{M}}(\Gamma \ | \ \mathcal{L})=1$.
	\end{description}
\end{definition}

Besides the practicality in monitoring applications, condition PAC3 also adds the counterfactual idea to the definition.
Since almost all executions in $\mathcal{L}$ have a prefix in $\Gamma$,
the effect  occurs with probability $0$ if the cause was not observed.
Condition PAC1\textsuperscript{p} is a third variant of the probability-raising requirement that compares 
the probability of the effect after observing an element of the cause to a specific threshold value $p$ rather than the overall 
probability $\Pr_{\mathcal{M}}(\mathcal{L})$. 
In case $p>\Pr_{\mathcal{M}}(\mathcal{L})$, this variant implies PAC1 and PAC1\textsuperscript{loc}.

For $\omega$-regular $\mathcal{L}$ there always exist $p$-causes for any $p\in (0,1]$. 
The reason is that then almost all paths in $\mathcal{L}$ have a prefix $\pi$ with $\Pr_{\mathcal{M}}(\mathcal{L}\mid \pi)=1$. 
Choosing the shortest of such prefixes in accordance with condition PAC2 yields a $1$-cause and hence a $p$-cause for any $p$.
For $p<1$, there are multiple $p$-causes in general. In \cite{BaierFJPZ21}, the problem to  
find cost-optimal $p$-causes with respect to a variety of cost measures is addressed.

The relationship between the different notions of causes is summarized in the following proposition. 
It is a direct consequence of the implications between the axioms used in the definitions that have been discussed so far.

\begin{proposition}
Let $\mathcal{M}$ be a DTMC with state space $S$, $\Gamma\subseteq S^+$, and 
let $\mathcal{L}\subseteq S^\omega$ be a measurable set of paths. Then the following statements hold:
\begin{enumerate}
\item If $\Gamma$ is a $p$-cause for $\mathcal{L}$ for some $p>\Pr_{\mathcal{M}}(\mathcal{L})$, then 
$\Gamma$ is also a local PR-cause  for $\mathcal{L}$.
\item If $\Gamma$ is a local PR-cause for $\mathcal{L}$, then $\Gamma$ is also a global PR-cause for $\mathcal{L}$. 
\item If $\Gamma$ is a singleton, then $\Gamma$ is a local PR-cause for $\mathcal{L}$ iff $\Gamma$ is a global PR-cause for $\mathcal{L}$.
\end{enumerate}
\end{proposition}

The probabilistic notions of causality discussed in this section naturally constitute forward-looking notions: 
the probability-raising principle  inherently addresses the behavior of a system across multiple executions, and 
causes are prone to exhibiting a predictive character. 
These notions can be useful in inferring causal dependencies in data series  \cite{KleinbergM2009,Kleinberg2011,HuangK15} and predicting undesirable behavior of reactive systems through runtime monitoring \cite{BaierFJPZ21}. 
Nevertheless, as far as formal probabilistic models are concerned, a comprehensive study of cause-effect relationships is still at the beginning.

\section{Concluding Remarks}

This article gave an overview of recent trends in causality-based reasoning
in the verification context. The focus of this article was on concepts that aim to explicate \emph{why} a system exhibits a specific observable behavior and to which degree individual agents of a system can be held \emph{responsible} for it. For non-probabilistic formal models, concepts of causation have been introduced in multiple facets and examined for manifold applications. 
To increase the power of causal inferences, a more systematic study relating forward and backward notions of causality would be highly beneficial.

Compared to the non-probabilistic setting, research on probabilistic causation in stochastic operational models is still in its infancy. While the techniques presented here are limited to purely probabilistic models (Markov chains), an examination of causality in probabilistic models with nondeterminism (Markov decision processes) is largely open. A first step in this direction is a formalization of action causes as a hyperproperty in Markov decision processes \cite{DimFinkbeinerTorfah-probHyper-MDP-ATVA2020}.
  Another important future direction is to reason about cause-effect relationships in hidden Markovian models where states (and events) are not fully observable.

Another research strand not covered in this article are causality-based verification techniques (see, e.g., \cite{KupriyanovF13,KupriyanovF14}) that rely on the successive identification of cause-effect relationships between events to generate a causality-based proof for the satisfaction or violation of a system property. Along these lines, the work \cite{BaierCFFJS21} presents a causality-based technique for solving symbolically expressed, infinite-state two-player reachability games. Applying this paradigm also in a probabilistic setting is a promising direction of study.

\bibliography{refs-final}

\end{document}